\documentclass[preprint,12pt]{aastex}
\usepackage{booktabs}
\usepackage{graphicx}
\usepackage{subfigure}
\usepackage{amsmath,amssymb,latexsym,graphics,epsfig,exscale,graphicx}
\usepackage{url}
\begin{document}

\title{High Resolution 8 mm and 1 cm Polarization of IRAS 4A from the VLA Nascent Disk and Multiplicity (VANDAM) Survey}
%
%
\author{Erin G. Cox\altaffilmark{1}, Robert J. Harris\altaffilmark{1}, Leslie W. Looney\altaffilmark{1},  Dominique M. Segura-Cox\altaffilmark{1},
John Tobin\altaffilmark{2}, Zhi-Yun Li\altaffilmark{3}, {\L}ukasz Tychoniec\altaffilmark{4}, Claire J. Chandler\altaffilmark{5}, Michael M. Dunham\altaffilmark{6}, Kaitlin Kratter\altaffilmark{7}, 
Carl Melis\altaffilmark{8}, Laura M. Perez\altaffilmark{5}, Sarah I. Sadavoy\altaffilmark{9}}
\altaffiltext{1}{\itshape Department of Astronomy, University of Illinois
at Urbana-Champaign, Urbana, IL 61801, USA;
egcox2@illinois.edu} 
\altaffiltext{2}{\itshape Leiden Observatory,  Leiden University, P.O. Box 9513, 2000-RA Leiden, The Netherlands}
\altaffiltext{3}{\itshape Department of Astronomy, University of Virginia, Charlottesville, VA 22903, USA}
\altaffiltext{4}{\itshape Astronomical Observatory Institute, Faculty of Physics, A. Mickiewicz University, S{\l}oneczna 36, PL-60-268 Pozna\'n, Poland}
\altaffiltext{5}{\itshape National Radio Astronomy Observatory, Socorro, NM 87801, USA}
\altaffiltext{6}{\itshape Harvard-Smithsonian Center for Astrophysics, Cambridge, MA 02138, USA}
\altaffiltext{7}{\itshape Steward Observatory, University of Arizona, Tucson, AZ 85721, USA}
\altaffiltext{8}{\itshape Center for Astrophysics and Space Sciences, University of California, San Diego, CA 92093, USA}
\altaffiltext{9}{\itshape Max-Planck-Institut f$\ddot{u}$r Astronomie, D-69117 Heidelberg, Germany}
%
%
%
%
\begin{abstract}
Magnetic fields can regulate disk formation, accretion and jet launching. Until recently, it has been 
difficult to obtain high resolution observations of the magnetic fields of the youngest protostars in the critical
region near the protostar.  The VANDAM survey is observing all known protostars in the
Perseus Molecular Cloud. Here we present the polarization data of IRAS 4A. We find that with $\sim$ 0.2$\arcsec$ (50 AU) resolution at $\lambda$ = 8.1 and 10.3 mm, the inferred magnetic field is consistent with
a circular morphology, in marked contrast with the hourglass morphology seen on larger scales.
This morphology is consistent with frozen-in field lines that were dragged in by rotating material entering the infall
region. The field morphology is reminiscent of rotating circumstellar material near the protostar. This is the first polarization detection of a protostar at these wavelengths. We conclude from our observations that the dust emission is optically thin with $\beta$ $\sim$ 1.3, 
suggesting that mm/cm-sized grains have grown and survived in the short lifetime of the protostar.
\end{abstract}

\keywords{circumstellar matter--- stars: formation--- stars: magnetic field}
\maketitle
%
%
%
%
\section{Introduction}  \label{sec:intro} 
Magnetic fields play a crucial role in star formation. On the large (cloud size) scale, they regulate
collapse through both ambipolar diffusion--enabling collapse--and magnetic pressure support--hindering 
collapse \citep{sh87}. These magnetic fields are responsible for funneling material from large
scale down to smaller (disk size) scales (Li et al. 2014). At these smaller scales, the magnetic field regulates disk accretion,
wind launching, outflows and jets \citep[e.g.,][]{mc07}. Finally, the magnetic field dictates whether proto-planetary disks 
can form and how they accrete. The small scale effects may have the most 
dramatic consequences for star formation and subsequently planet formation. In the ideal MHD limit, if the alignment 
of the field is parallel to the rotation axis of the infalling envelope, then magnetic braking occurs, hindering 
large disk growth at the earliest \citep[e.g., Class 0 object,][]{an00} protostellar stage \citep[e.g.,][]{ml08,hf08,li11}.
 However, if the magnetic field 
is perpendicular, or misaligned, to the rotation axis of the envelope, then a large disk can grow in a Class 0 
object \citep[R $\sim$ 100 AU; e.g.,][]{jo12}. Simulations including non-ideal MHD effects show that
small (R $\sim$ 10 AU) disks can indeed grow in Class 0 objects, but do not reach larger radii until they are more evolved, Class II objects \citep[]{db10,to15,in15,li14}.
This ambiguity in the importance of magnetic fields makes observations of the youngest sources vital to truly understanding the role magnetic fields play in star formation. 

Recent high resolution observations of Class 0 sources have provided important observational constraints
for theory. The Telescope Array Doing Polarization
(TADPOL) survey was a flux-limited, 1.3 mm polarization survey of 30 cores and 8 star forming 
regions to characterize their magnetic fields on 1000 AU size scales \citep[]{hu13,hu14}.
TADPOL found that, on average, the inferred magnetic field is misaligned with the outflow 
direction (used as a proxy for disk rotation axis), in contrast to the fact that they can appear aligned 
at the 10000 AU scale \citep{ch13}. These results suggest that magnetic field 
morphology can change drastically from large to small scales and is potentially significant for 
understanding the role of magnetic fields in the star formation process. Similarily, very high resolution 
observations have confirmed that for some of the Keplerian Class 0 disks that are known, the magnetic 
field orientations and disk planes are perpendicular to outflow directions consistent with misaligned fields, 
enabling disk formation (VLA 1623, \citealt{mu13}; and L1527, \citealt{to12,to13,sc15}). We used the Karl G. Jansky Very Large
Array (VLA) to conduct the VLA Nascent 
Disk and Multiplicity Survey (VANDAM)--a high resolution survey of continuum emission at $\lambda =$ 8.1 mm, 10.3 mm,
4.1 cm and 6.4 cm from all known protostars in the Perseus region \citep[]{tob15} to determine structure, multiplicity and polarization properties of these young protostars. One
of our sources, NGC1333 IRAS 4A (hereafter IRAS 4A), has long been an ideal candidate for polarization observations at long wavelengths
because of its high degree of polarization at millimeter wavelengths \citep{gi06}.
 
Located in the Perseus Molecular Cloud \citep[d $\sim$ 230 pc,][]{hi08,hi11}, IRAS 4A is a nearby, Class 0 
protostar \citep[as seen by the SED in][]{sa91,an93}. This system is actually made up of two protostars in
a circumbinary envelope \citep[e.g.,][]{la95,lo00}. Multiple observations of 
IRAS 4A have shown a well-defined outflow arising from the binary system. \citet{sa15p} resolves
both outflows from the two sources. In addition to having multiple observations, IRAS 4A is incredibly bright in both the mm/sub-mm bands. 

In this letter, we present unprecedented high-resolution polarization observations of IRAS 4A 
from the VLA at 8.1 mm and 10.3 mm wavelengths. This is first ever polarization detection at these wavelengths
in a protostar. We show the observed morphology of the inferred
magnetic field and attempt to link it to the morphology seen at larger scales. We posit reasons
why our data differ from earlier observations, and we use simple modeling to interpret observations.

\section{Observations}  \label{sec:obs}
Observations of IRAS 4A were taken as part of the VANDAM 
survey. IRAS 4A was observed in full Stokes using the B-array of the VLA in Ka-band on 21 October 
2013 with baselines ranging from 0.21 to 11.1 km. These observations of IRAS 4A included two other nearby science targets.
All science scans were interleaved with scans of the complex gain calibrator J0336+3218. We used
3C48 as our flux and polarization position angle calibrator, and 3C84 was used as our bandpass and leakage calibrator. We observed a second angle calibrator, 3C138, to check measurement of the polarization angle. IRAS 4A was also observed in A-array in C-band ($\lambda$ = 4.1 and 6.4 cm) on 16 March 2014, using 3C48 as the bandpass
and flux calibrator \citep[see for further details,][]{tob15}.
	
The correlator setup for Ka-band had two basebands both with 4 GHz of bandwidth. Each of these basebands
was divided into 32 spectral windows, with each window consisting of 64 channels. The spectral windows
had a bandwidth of 128 MHz. Basebands were centered at $\lambda$=8.1 mm and 10.3 mm. All four correlations
were recorded to permit full Stokes imaging.
	
We first calibrated the data collected for IRAS 4A in \texttt{CASA} for delay, bandpass and complex gain variations in the
antennas. After the initial calibrations were done, we performed polarization calibrations that
solved for leakage, cross-hand delay and R-L phase differences across the array. Once these calibrations
were done, the data were spectrally averaged into two data sets of 32 channels each, one for each of the two basebands.
We then \texttt{SPLIT} out the science targets from the data set. The amplitude calibration
for Ka-band has $\sim$10\% uncertainty, and for C-band there is $\sim$5\% uncertainty. For the Stokes calibration, we only
consider statistical uncertanties.
	
Once the calibration was done, we used \texttt{CLEAN} to deconvolve the images. Natural weighting 
was used to produce full Stokes images of Ka-band emission, maximizing sensitivity to polarization. As a cross-check we imaged the other two science targets and gain calibrator. We found no polarization in the two science targets down to $\lesssim$5.5\%. Our gain calibrator was found to be polarized at $\sim$1\%, as expected. The synthesized beam of
the observations was 0.224$\arcsec$ $\times$ 0.199$\arcsec$ with a position angle of -81.1${^\circ}$ at 8.1 mm. At 10.3 mm, our
synthesized beam was 0.25$\arcsec$ $\times$ 0.25$\arcsec$ with a position angle of -80.1${^\circ}$. To optimize the tradeoff between sensitivity and resolution, Briggs weighting with a robust parameter equal to 0.25 was used to image the C-band emission. 
The C-band data had a
synthesized beam of 0.42$\arcsec$ $\times$ 0.35$\arcsec$ with a position angle of 84.4${^\circ}$ for 6.4 cm and a beam size of
0.27$\arcsec$ $\times$ 0.23$\arcsec$ with a position angle of 85.5${^\circ}$ for 4.1 cm.	
\section{Results}  \label{sec:results}
\begin{figure*}[ht]
\includegraphics[width=1.1\textwidth]{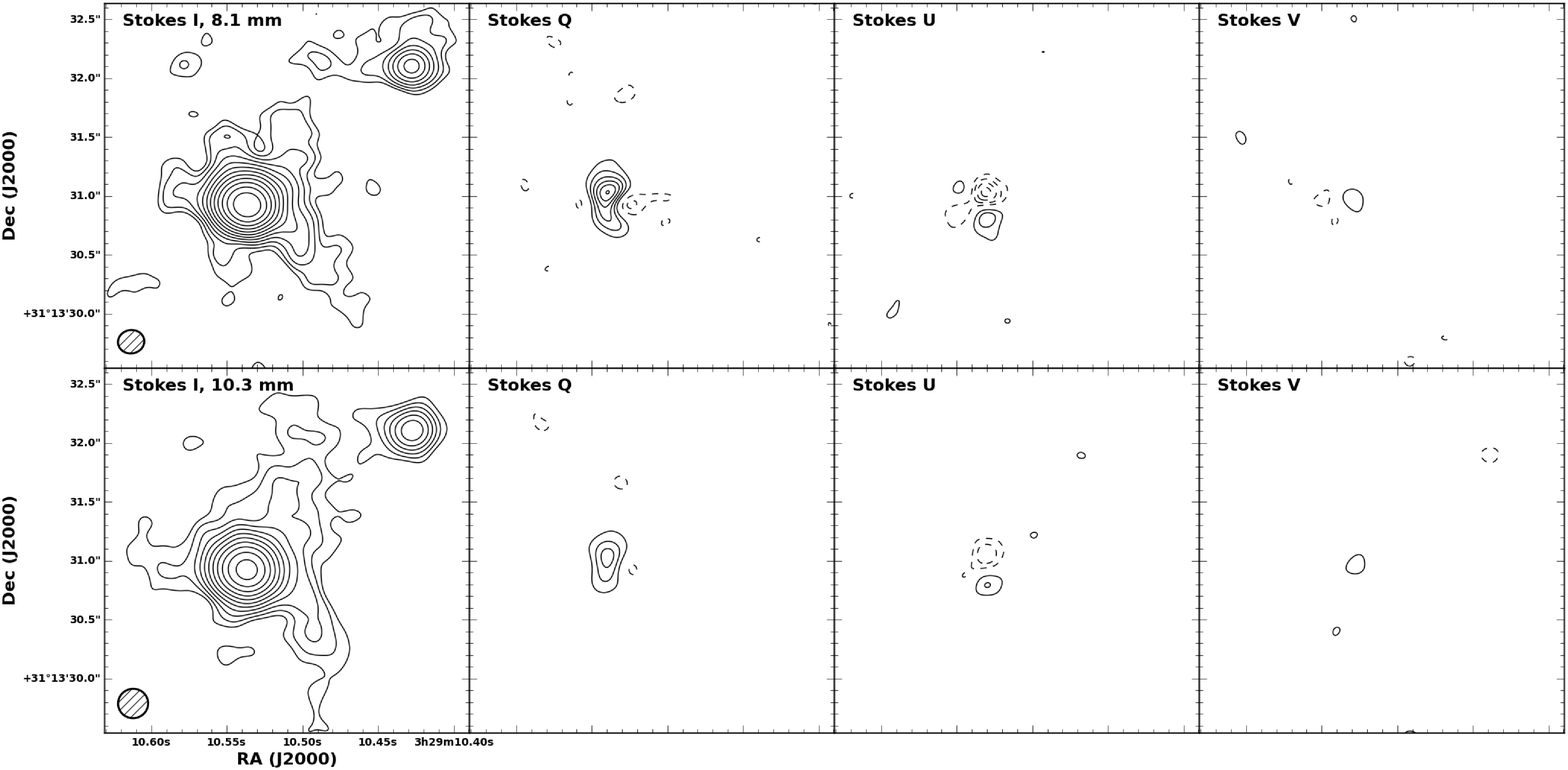}
\small\caption{\small \setlength{\baselineskip}{0.85\baselineskip}
Images of the 8.1 mm (upper panel) and 10.3 mm (lower panel) continuum emission for IRAS 4A in each of the four Stokes parameters
taken using the VLA B-array. Contours start at 3$\sigma$. For Stokes I contours are separated by a factor of $\sqrt{2}$ while
for QUV they increase linearly in steps of 3$\sigma$. At 8.1 mm, the 1$\sigma$ noise levels are 23 $\mu$Jy/beam for Stokes I and
$\sim$ 14 $\mu$Jy/beam for Stokes QUV. At 10.3 mm, the 1$\sigma$ noise levels are 25 $\mu$Jy/beam for Stokes I and $\sim$ 13 $\mu$Jy/beam for Stokes
QUV. Dashed contours indicate negative values. The synthesized beam is shown in bottom left corner.
}
\vspace{-0.2in}
\end{figure*}

\begin{table}
\begin{center}
\begin{tabular}{|| c c c c c c c ||}\toprule
Source & $\lambda$ (mm) & \multicolumn{2}{c}{Stokes I} & Stokes Q  & Stokes U & Stokes V \\
  & & peak   &  integrated & peak & peak & peak \\
 & & ($\mu$Jy/beam) & ($\mu$Jy) & ($\mu$Jy/beam) & ($\mu$Jy/beam) & ($\mu$Jy/beam) \\
\hline
\hline
 1 (SE) & 8.1 & 4151$\pm$54  &  9770$\pm$130 & 220$\pm$28 & -197$\pm$14 & $\lesssim$40.3 \\
  & & & & -91$\pm$23 & 117$\pm$14 & \\
 1 (SE) & 10.3 & 2759$\pm$29 &  5488$\pm$58 & 137$\pm$14 & -112.6$\pm$14 &$\lesssim$14  \\
  & & & & -50.8$\pm$3.9 & 79.5$\pm$14&  \\
  1 (SE) & 41 & 92$\pm$4 & 117$\pm$8 & & & \\
  1 (SE) & 64 & 47$\pm$5 & 54$\pm$9 & & & \\ 
 \hline
 2 (NW) & 8.1 & 601$\pm$25&  955$\pm$25  &$<$40.0 & $<$40.4 &$<$40.3 \\
 2 (NW) & 10.3 & 545$\pm$25 & 695$\pm$28 & $<$40.5& $<$37.7&$<$13.5  \\
 2 (NW) & 41 & 130$\pm$4 & 138$\pm$6 & & & \\
 2 (NW) & 64 & 103$\pm$5 & 102$\pm$6 & & & \\
\hline
\end{tabular}
\end{center}
\caption{Peak and integrated flux densities in each of the four Stokes parameters for both the southern source IRAS4 A1 (denoted by 1 here) and the northern source IRAS4 A2 (denoted by 2 here) For Stokes Q and U, we fit both the positive and negative point-like components seen in the vicinity of IRAS4 A1 (see Fig. 1). Errors on parameters reported are derived from the formal errors returned from the Gaussian fits. In cases where the map noise is larger, then the true uncertainty given by the map rms is reported. Polarization was performed only for Ka-band and not for C-band.}
\end{table}

In Figure 1, we present two sets of images of IRAS 4A in full Stokes radio continuum emission. At
both 8.1 and 10.3 mm in Stokes I, there are two bright sources (IRAS 4A1 and IRAS 4A2, 
in the south and north, respectively). These two compact sources are the radio counterparts to 
the mm-wave sources first resolved at 0.84 mm \citep{la95} and at 3 mm \citep{lo00}.
We detect two bright compact protostars surrounded by some extended emission. In Table 1, we present
peak and integrated flux densities for each component in the system. The spectral index between 8.1 and 10.3 mm
for IRAS 4A1 is $\sim$2.2$\pm$0.1, consistent with being dominated by dust emission, while that between 4.1 and
6.4 cm is $\sim$1.7$\pm$0.4. For IRAS 4A2, the spectral index between 8.1 and 10.3 mm is $\sim$1.22$\pm$0.1
and that between 4.1 and 6.4 cm is $\sim$-0.68$\pm$0.16. The flatter slope for IRAS 4A2 is consistent with a
substantial contribution from free-free emission. At 4.1 and 6.4 cm the protostars are detected as unresolved 
point sources with little to no extended emission.

\begin{figure*}[htp]
  \centering
  \begin{tabular}{cc}
   \includegraphics[width=70mm]{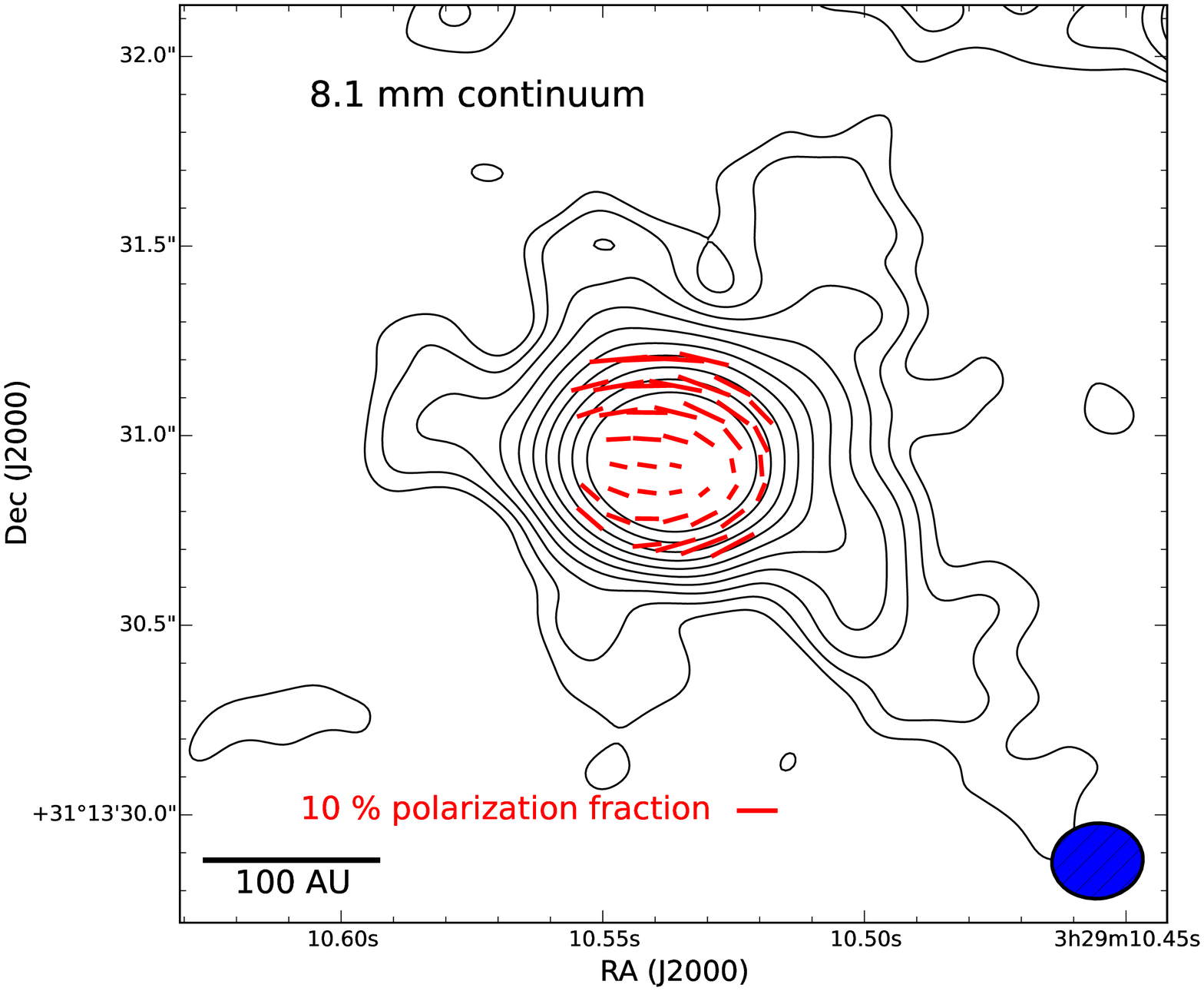}&
   \includegraphics[width=70mm]{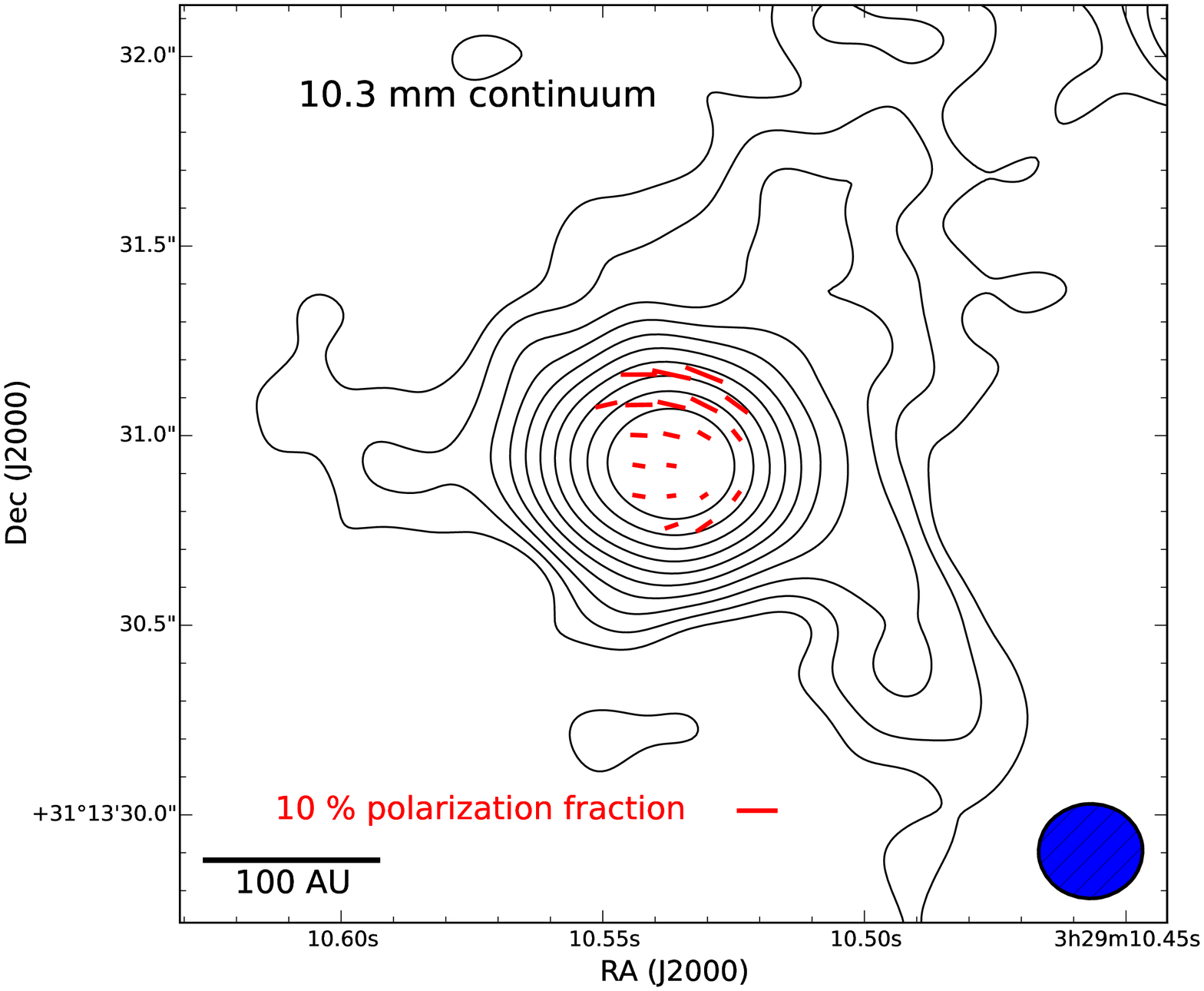}
    \end{tabular}
 \caption{Left panel, inferred magnetic field from polarization at 8.1 mm. Right panel, inferred magnetic field at 10.3 mm. Contour spacing is the same as in the Stokes I panel in Figure 1. The polarization maps were made using a 3$\sigma$ mask in I.}
\label{Fig:Magnetic Field}
\end{figure*}

We present maps of linear polarization intensity and inferred magnetic field direction in Figure 2. To obtain the polarization intensity map, we used the formula $P = \sqrt{Q^{2}+U^{2}}$ and to get the fractional levels, we divided this number by the total intensity.  We only detect polarized emission towards IRAS 4A1 (SE source) in this map: 
there is no polarization detected towards IRAS 4A2 at a 3$\sigma$ level of $\sim$60 $\mu$Jy.
The polarized emission is more compact than the total intensity but is still resolved. For the magnetic field maps, we first derived the polarization angle by using the formula $\alpha$ = 1/2arctan(U/Q). We rotated the angle 
by 90$^{\circ}$ to form the inferred magnetic field map. These maps show a magnetic field consistent with a circular morphology near the protostar.

To verify that the polarized emission is most likely tracing the magnetic field, we can confirm that the emission is dust dominated through an examination of the spectrum from 8 mm to 64 mm. We use the spectrum to estimate 
the contamination of the dust emission at Ka-band by free-free emission. In principle, free-free emission can have any spectral index between -0.1 and 2, depending on optical depth and source geometry \citep[e.g.,][]{pa75,re86}. Assuming
that all of the flux between 4.1 cm and 6.4 cm is free-free emission, we computed the best fit power law. We then
extrapolated the free-free fit to 8.1 mm and 10.3 mm to determine the relative contribution of dust and free-free
emission at 8.1 mm and 10.3 mm. After adding in 1.3 mm \citep{jo07} and 2.7 mm \citep{lo00} flux values, we fit the spectrum with a power law (seen in Figure 3 for IRAS 4A1). This gives a dust emission spectral index $\alpha$$\sim$3.2$\pm$0.07 for IRAS 4A1 and $\alpha$$\sim$3.4$\pm$0.2 for IRAS 4A2. However, the shorter wavelength observations have lower resolution and less spatial filtering as compared to our data; thus the shorter wavelength data will tend to raise the measured spectral index due to including more flux from larger scales. We find that the fraction of emission from dust at 8.1 mm is $\sim$0.83 and at 10.3 mm it is $\sim$0.74 for
IRAS 4A1, while for IRAS 4A2 at 8.1 mm it is $\sim$0.59 and at 10.3 it is $\sim$0.46.

\begin{figure*}[h]
\includegraphics[width=0.9\textwidth]{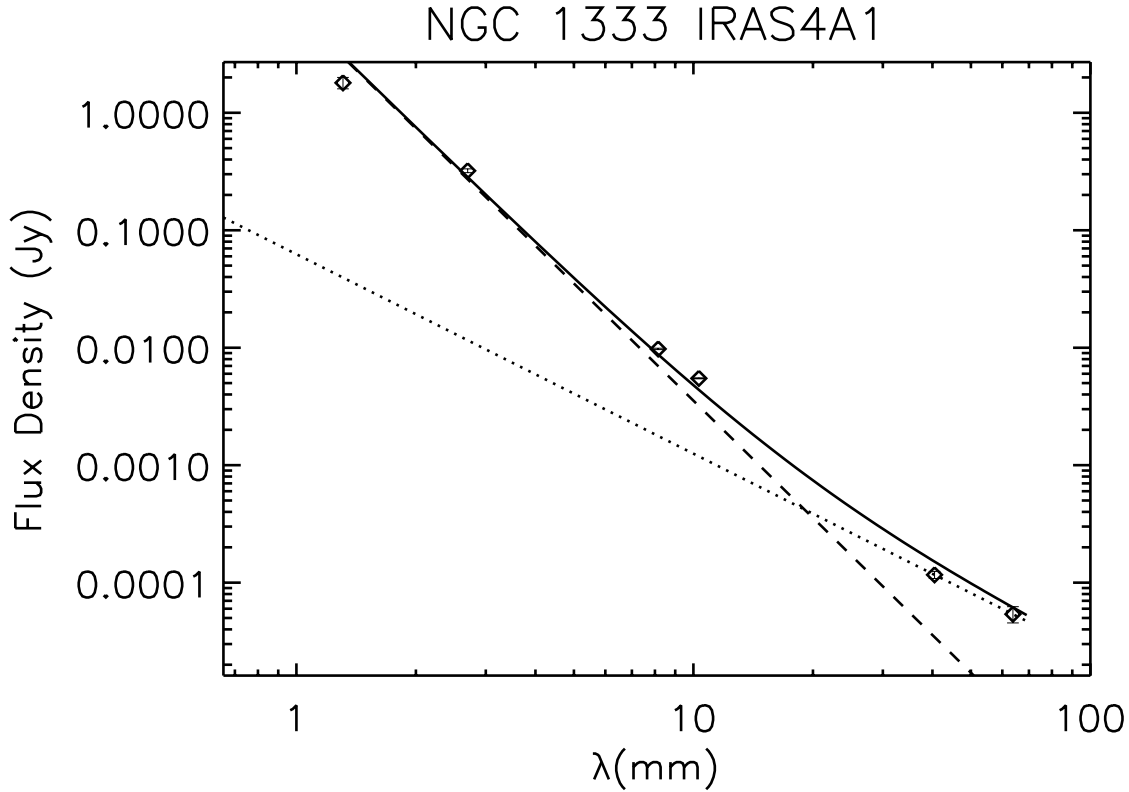}
\small\caption{\small \setlength{\baselineskip}{0.85\baselineskip}
Spectrum of IRAS 4A1. The 1.3 mm flux from \citet{jo07} is 1800 $\pm$ 600 mJy IRAS 4A1, and the 2.7 mm flux from \citet{lo00} is 324.1$\pm$ 12.0 mJy. The dotted line is the estimated free-free
emission and the dashed line is the dust emission. The solid line is the sum of the two.
}
\vspace{-0.1in}
\end{figure*}

Since the continuum emission is dominated by dust, we can estimate the mass of the inner envelope and disk. We assumed 
optically thin dust emission in the Rayleigh-Jeans limit \citep{hi83}:
\begin{equation}
M=\frac{d^2F_{\nu}}{B(T_d)\kappa_{\nu}}
\end{equation}
where $M$ is the total circumstellar mass, $d$ is the distance, $F_{\nu}$ is the integrated flux, $B(T_d$) is the Planck function
evaluated at our assumed dust temperature and $\kappa_\nu$ is our assumed total (gas + dust) opacity. We assume 
a distance of 230 pc and a dust temperature of 30 K. Our total opacity is given by \citep{oh93}:
\begin{equation}
\kappa_{total}=0.00899\left(\frac{1.3~ \text{mm}}{\lambda}\right)^\beta\text{cm$^2$g$^{-1}$}
\end{equation} 
assuming a 100:1 gas to dust ratio, where $\beta$$\sim$1.3. We determine the mass
of the circumstellar material surrounding IRAS 4A1 to be $\sim$2.3M$_\odot$. While
this is estimate is quite high, the calculation is subject to considerable uncertainty given that the appropriate dust opacity 
millimeter/centimeter bands is not well-known.
\section{Discussion \& Conclusions}  \label{sec:disc}
In the above section, we presented $\sim$0.2$\arcsec$ resolution images of the 8.1 and 10.3 mm continuum
in full Stokes, as well as the inferred magnetic field of IRAS 4A1. We see that in the plane of
the sky, the magnetic field morphology is circular at both wavelengths. This might be expected for a magnetized rotating disk 
\citep[e.g., the simple, face-on disk model in][]{ra14}. While there is no confirmed disk surrounding IRAS 4A1 yet, kinematic observations confirm a velocity gradient across the source, consistent with rotation \citep{ye15}, and our observations reveal
what looks to be a toroidal field that is being wrapped by a rotating (close to face-on) disk or inner envelope \citep{hc09,ka14}. Such a polarization signal is consistent with the expectation that magnetic fields misaligned with rotation axes do not inhibit disk growth and with several observational results that support the theoretical expectation (e.g., L1527, \citealt{sc15}; HL Tau, \citealt{st14}). While our observations are unable to distinguish between a disk and the inner envelope, the magnetic field 
morphology suggests it originates from circumstellar material rotating close to the central protostar. 

Our observations contrast sharply with the hourglass field morphology seen in $\sim$1$\arcsec$ resolution SMA observations of the 850 $\mu$m continuum \citep{gi06,gi08}. There are several possible explinations for this discrepancy. First, the spatial filtering
of the SMA observations in \citet{gi06} and of our observations with the VLA, are quite different: we expect
to resolve out much of the large scale structure of the source. Relatedly, the spatial
resolution of the SMA was $\sim$250 AU and probes envelope-sized structure, whereas our VLA data
probes much smaller ($\sim$50 AU), disk size scales. As material falls from the envelope to a nascent disk, it is likely
to drag frozen-in magnetic field lines with it. As it does, the magnetic field will change in morphology from the
envelope to the disk \citep{li14}; so we are detecting this region. If the magnetic field then mirrors the disk/inner
envelope rotation, then this could cause the difference in morphologies. Future lower-resolution VLA observations taken in either the C or D configurations may show a smooth transition from the polarization structure probed by the B array data to that probed by the SMA data at the shorter millimeter wavelengths \citep{fr11}. In fact, we produced images giving more weight to shorter \textit{uv}-spacings, yielding a resolution of $\sim$0.5$\arcsec$ and we saw morphology which more closely resembles that seen in lower resolution observations.

\begin{figure*}[htp]
  \centering
  \begin{tabular}{cc}
   \includegraphics[width=70mm]{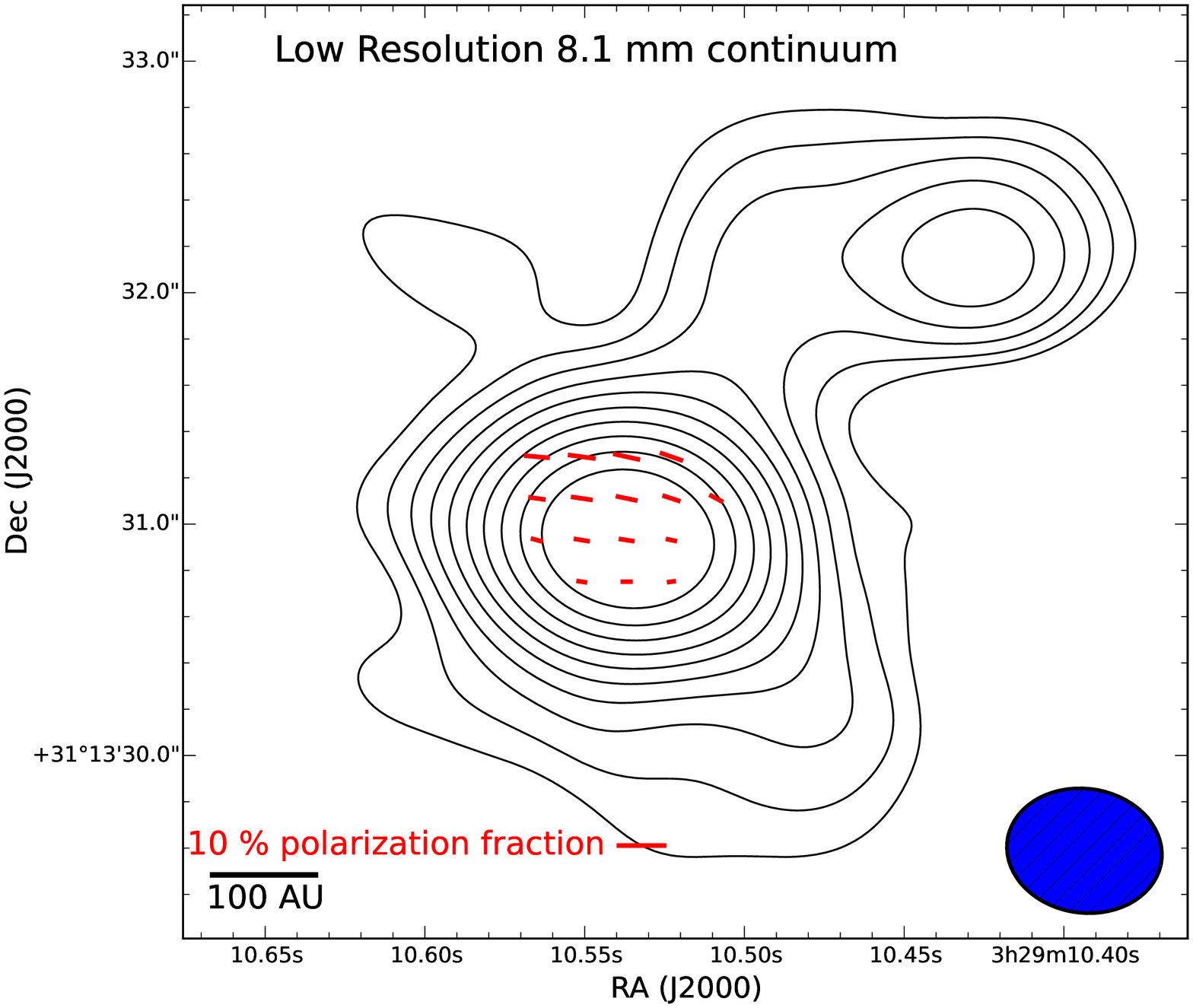}&
   \includegraphics[width=70mm]{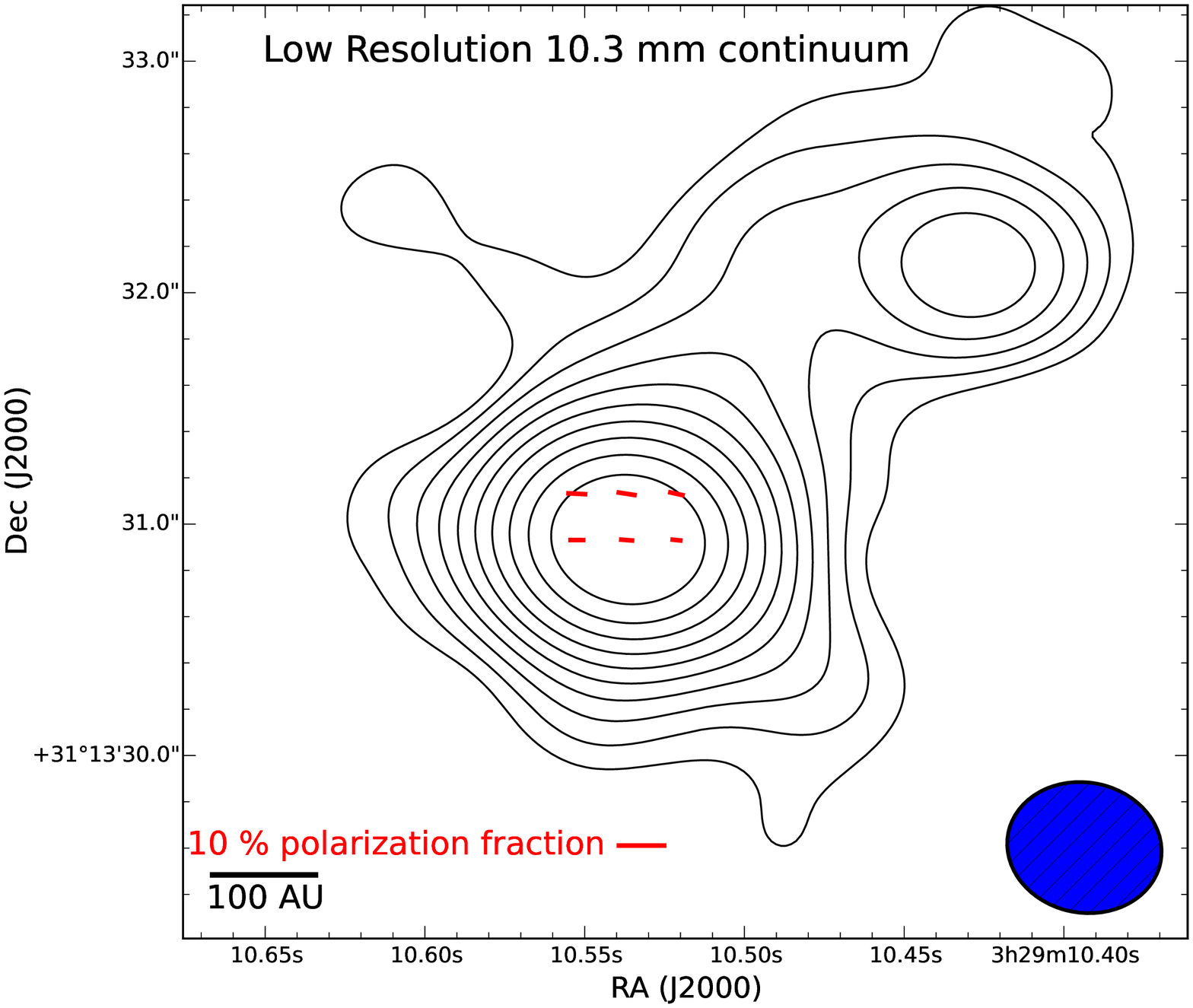}
    \end{tabular}
 \caption{Low resolution maps of Stokes I and polarized emission as made from imaging the data with a Gaussian taper to produce a resolution of 0.63$\arcsec$ $\times$ 0.54$\arcsec$ at 8.1 mm and 0.63$\arcsec$ $\times$ 0.56$\arcsec$ at 10.3 mm. The position angles are at 78.6${^\circ}$ and 73.1${^\circ}$ at 8.1 mm and 10.3 mm, respectively. The rms at 8.1 mm was 61 $\mu$Jy/beam and 41$\mu$Jy/beam at 10.3 mm. The contours indicate Stokes I, starting at 
 3$\sigma$ and increasing by $\sqrt{2}$. The magnetic field vectors show a much different morphology than what is seen at higher resolutions and also a much lower polarization percentage.}
\label{Fig:lowres Magnetic Field}
\end{figure*}

From our SED, we derived a value of $\alpha$ $\sim$ 2.2 which, when corrected for free-free contamination, corresponds to a
dust spectral index $\alpha_{\mathrm{dust}}$ $\sim$ 3.2 and a $\beta$ value of 1.2. Because these fit values incorporate the relatively lower-resolution short wavelength data,
they are upper limits, so $\alpha_{\mathrm{dust}}$ $\lesssim$ 3.2 and $\beta$ $\lesssim$ 1.2. Since the free-free source is weak, we can infer
that the centimeter emission is $\sim$80\% from dust emission. The polarization signal must be substantially due to dust emission
as well. Free-free emission is an unpolarized emission process; other than dust emission from grains aligned in a magnetic field, the other
potential process creating polarized emission could be synchrotron emission. However, the spectrum is inconsistent with synchrotron emission, which would be expected to have a negative spectral index at these frequencies. This flux level and $\beta$ value implies that large numbers of $\sim$ millimeter/centimeter-sized particles must have already grown and, also, survived in the 10 kyr age of IRAS 4A \citep{sa91}; similarly low values of $\beta$ have been found for other Class 0 sources, \citep[e.g., L1448 IRS 3, L1157][]{kw05,ch12,sch14,to13}. If our polarization
signal does indeed trace a Class 0 disk, this could have significant implications for models of grain-growth and
evolution within proto-planetary disks and how quickly grains can form in such an environment. 

The combination of previous SMA 850 $\mu$m polarization observations \citep{gi06} and our VLA 8.1 and 10.3 mm polarization observations provide insight into how the efficacy of grain alignment with magnetic field changes as a function of particle size.  Modeling by \citet{cl07} indicates that, at least for particles of size about a millimeter or smaller, the tendency of grains to align with the magnetic field decreases as a function of particle size. While this model has not been extrapolated to the $\gtrsim$ millimeter-sized particles that we know are contributing to the dust emission that we observe, it is commonly accepted that the polarization fraction does not increase with particle size past 1 mm. The observational counterpart to this theoretical prediction is a decreasing polarization fraction with wavelength. However, the SMA observations indicated a $\sim$few percent polarization towards the outskirts of the envelope and $\lesssim$2 percent polarization towards the central protostar (IRAS 4A1). In contrast, we find a remarkably high average polarization of 15\% at 8 mm and 18\% at 10 mm, with a peak fractional polarization of $\sim$20\%. While we described above a few potential explanations for the relatively low fractional polarization
of IRAS4A 1 at millimeter wavelengths, it is nonetheless striking that dust emission at this wavelength could be so highly polarized at centimeter wavelengths. Figure 1 in \citet{gi06},
shows a fairly low ($\lesssim$2\%) polarization in the same spatial area where we observe
$\sim$10-20\%. In lower resolution maps, we find that the beam averaging suppresses the polarization signal and at $\sim$0.5$\arcsec$ we find a maximum polarization of $\sim$4\%. This suggests that beam effects play a role in what the SMA is seeing, although optical depth effects could also be at work--as the 850 $\mu$m emission is likely to be much more optically thick than the 8.1 and 10.3 mm emission. 

This is the first time that polarization has been detected in a protostar at these wavelengths. Our results 
suggest that large grains may grow at early times. On the small scale, we observe a very different magnetic field morphology than on the large scale, showing frozen-in field lines are likely dragged in by infalling material. With our results, we have demonstrated the versatility of the VLA for detecting polarization on sub-100 AU scales.

The authors would like to thank the referee for their quick review and helpful comments, which helped improve the quality of this 
paper. The National Radio Astronomy Observatory is a facility of the National Science Foundation operated under cooperative agreement by Associated Universities, Inc. EGC, RJH, LWL and DMSC acknowledge support from NSF AST-1139950. JJT acknowledges support by grant 639.041.439 from the Netherlands Organisation for Scientific Research (NWO). ZYL is supported in part by NSF AST-1313083 and NASA NNX14AB38G. This research made use of APLpy, an open source plotting package for Python hosted at \url{http://aplpy.github.com}.

\acknowledgments

\end{document}